\documentclass[aps,prd,superscriptaddress,floatfix,showpacs]{revtex4}

\usepackage{verbatim}

\usepackage{amsmath}
\usepackage{amsfonts}
\usepackage{amssymb}
\usepackage{graphicx}
\usepackage{bm}
\usepackage{color}
\usepackage{epsf}
\usepackage{psfrag}
\newcommand{\beq}[1]{
\begin{equation}\label{#1}}
\newcommand{\eeq}{\end{equation}}
\newcommand{\bea}[1]{
\marginpar{\small\textsf{#1}}
\begin{eqnarray}\label{#1}}
\newcommand{\eea}{\end{eqnarray}}
\def\bea{\begin{eqnarray}}
\def\eea{\end{eqnarray}}
\def\beas{\begin{eqnarray*}}
\def\eeas{\end{eqnarray*}}
\def\beqas{\begin{eqnarray*}}
\def\eqas{\end{eqnarray*}}
\def\beq{\begin{equation}}
\def\eeq{\end{equation}}
\def\beqd{\begin{displaymath}}
\def\eeqd{\end{displaymath}}
\def\eqd{\end{displaymath}}

\def\slashchar#1{\setbox0=\hbox{$#1$}
   \dimen0=\wd0
   \setbox1=\hbox{/} \dimen1=\wd1
   \ifdim\dimen0>\dimen1
      \rlap{\hbox to \dimen0{\hfil/\hfil}}
      #1
   \else\begin{eqnarray}
      \rlap{\hbox to \dimen1{\hfil$#1$\hfil}}
      /
   \fi}

\begin{document}
\title
{Hard exclusive neutrino production of a  light meson }
\author{ B.~Pire}
\affiliation{ Centre de Physique Th\'eorique, \'Ecole Polytechnique,
CNRS, Universit\'e Paris-Saclay, 91128 Palaiseau,     France }

\author{ L.~Szymanowski}
\affiliation{ National Centre for Nuclear Research (NCBJ), 00-681 Warsaw, Poland}

\author{  J. Wagner}
\affiliation{ National Centre for Nuclear Research (NCBJ), 00-681 Warsaw, Poland}
\date{\today}
\begin{abstract}

\noindent
We update the leading order in $\alpha_s$ QCD amplitude for deep exclusive  neutrino and antineutrino production of a  light meson on an unpolarized nucleon. The factorization theorems of the collinear QCD approach allow us to write the amplitude as the convolution of generalized parton distributions (GPDs) and  perturbatively calculable coefficient functions. We study both the pseudoscalar meson and longitudinally polarized vector meson cases. It turns out that, contrarily to the electroproduction case, the leading twist scattering amplitudes for $\pi$ and $\rho_L$ productions are proportional to one another, which may serve as an interesting new test of the leading twist dominance of exclusive processes at medium scale. The dominance of the gluonic contribution to most cross sections  is stressed.
\end{abstract}
\pacs{13.15.+g, 12.38.Bx, 24.85.+p, 25.30.Pt}

\maketitle

\section{Introduction.}

Besides deeply virtual Compton scattering and deep exclusive meson leptoproduction \cite{DVCS}, deep exclusive neutrino production of a meson  \cite{weakGPD,PS,PSW} is a  way to access generalized parton distributions (GPDs) in the framework of collinear QCD factorization \cite{fact1,fact2,fact3}. Because previous studies \cite{Kopeliovich:2012dr} omitted the leading order gluon contributions, we update the predictions for light meson production cross sections.
We write the  scattering amplitude $W ~N \to M ~N'$   in the collinear QCD framework as a convolution of  leading twist quark  and gluon GPDs with a  coefficient function calculated in the collinear kinematics.

In this paper we consider the exclusive production of a pseudoscalar $M = \pi$ or longitudinally polarized \footnote{It is straightforward to see that the argument \cite{rhoT} for a zero leading twist amplitude for the production of a transversally polarized vector meson applies as well to neutrino production as to leptoproduction.}  vector $M = \rho_L$ meson through the reactions :
\begin{eqnarray}
\nu_l (k)N(p_1) &\to& l^- (k') M (p_M)N'(p_2) \,,\\
 \bar\nu_l (k) N(p_1) &\to& l^+ (k') M(p_M) N' (p_2)\,,
\end{eqnarray}
in the kinematical domain where collinear factorization  leads to a description of the scattering amplitude 
in terms of nucleon GPDs and the $M$ meson leading twist distribution amplitude, with the hard subprocesses described by the  handbag Feynman diagrams of Fig. 1 convoluted with chiral-even quark GPDs, and the corresponding ones of Fig. 2 convoluted with gluon GPDs.

%%%%%%%%%%%%%%%%%%%%%%%%%%%%%%%%%%%%%%%%%%%%%%%%%%%%%
\begin{figure}
\includegraphics[width=0.8\textwidth]{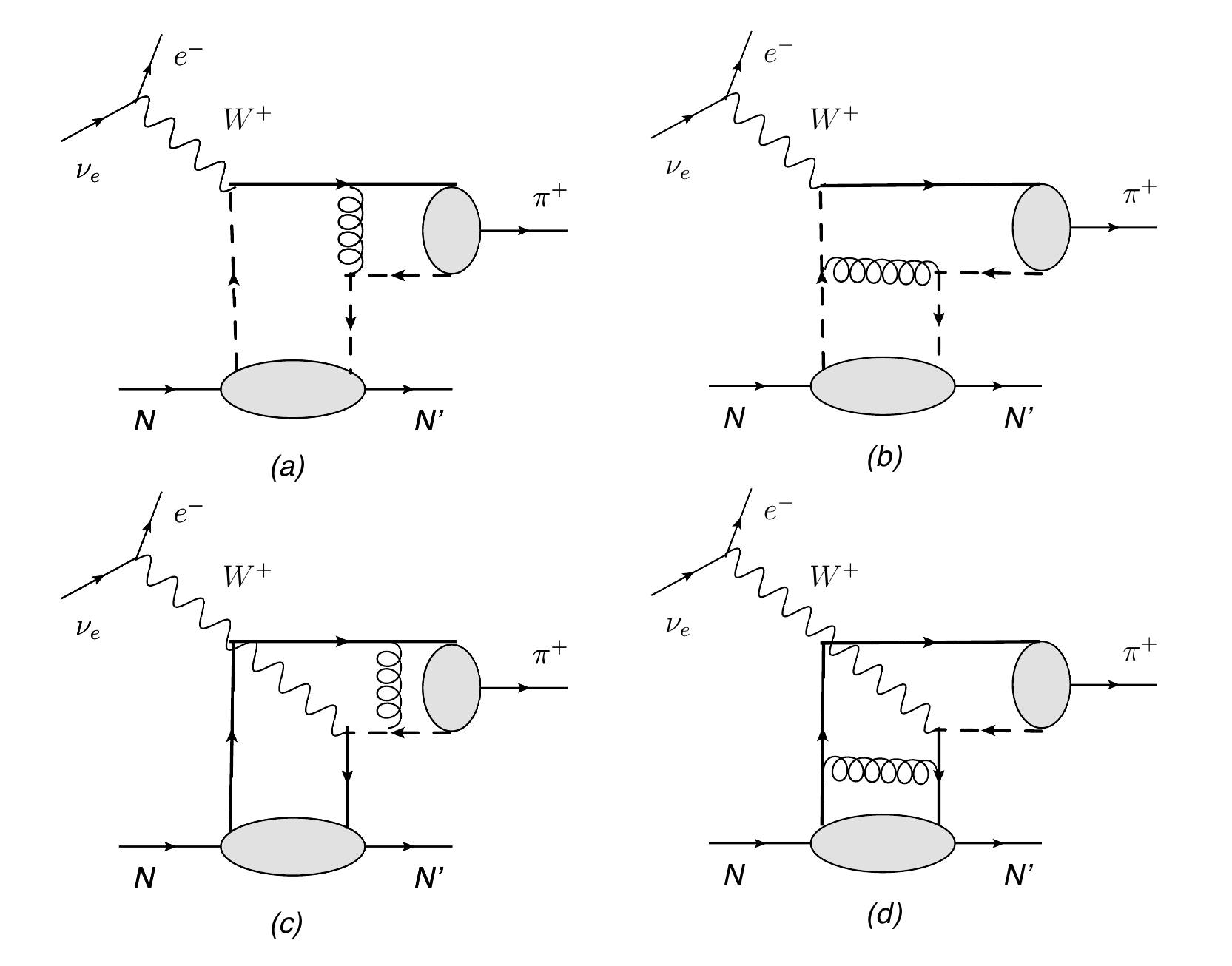}
%\vspace{1cm}
\caption{Feynman diagrams for the factorized  amplitude for the $ \nu_l N \to l^-  \pi^+ N'$  process involving the quark GPDs. The solid line represents the $u-$quark, the dashed line represents the $d-$quark.}
   \label{Fig1}
\end{figure}

%%%%%%%%%%%%%%%%%%%%%%%%%%%%%%%%%%%%%%%%%%%%%%%%%%%%%
\begin{figure}
\includegraphics[width=0.8\textwidth]{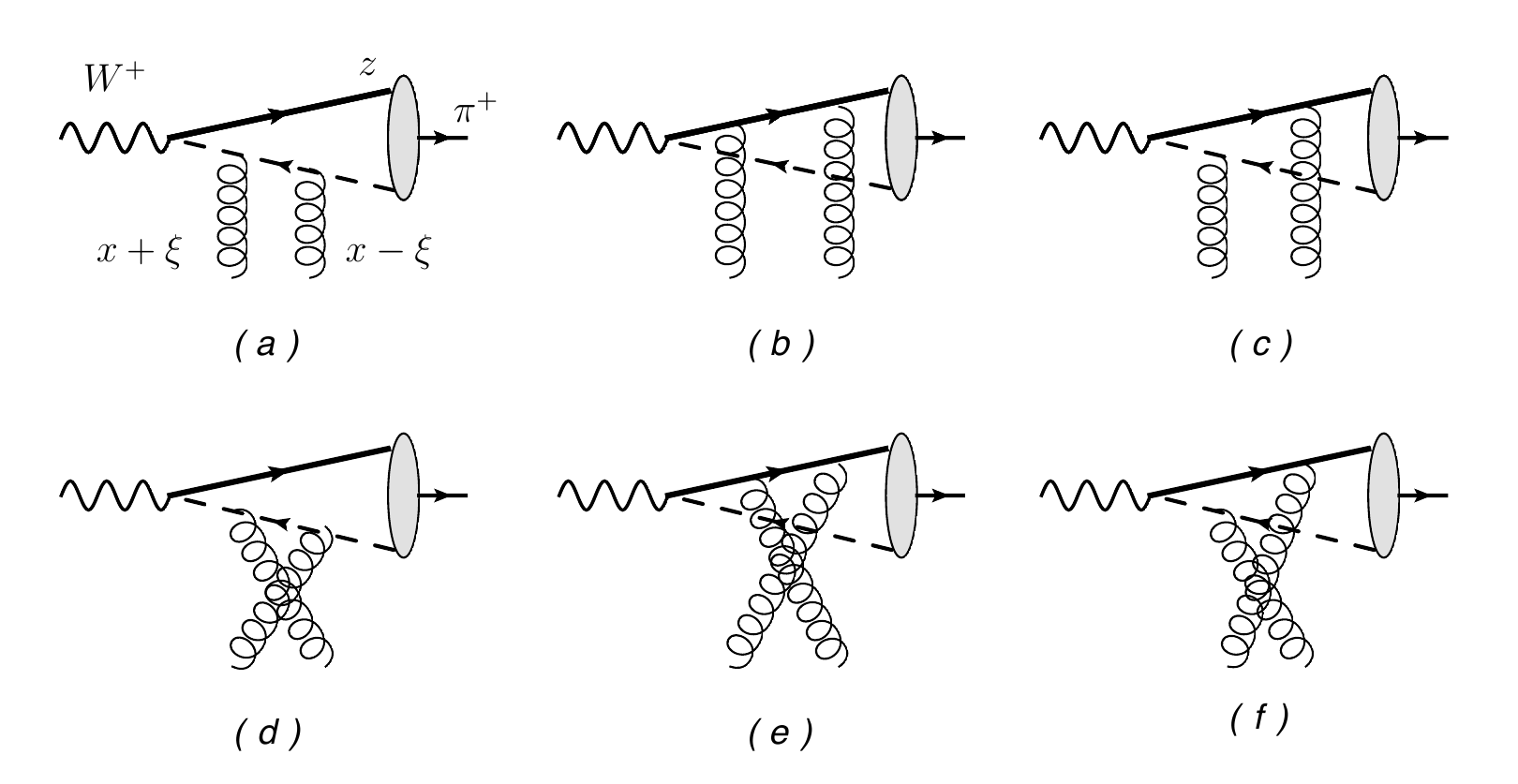}
%\vspace{1cm}
\caption{Feynman diagrams for the factorized  amplitude for the $W^+ N \to M^+ N'$ process involving the gluon GPDs. The solid line represents the $u-$quark, the dashed line represents the $d-$quark.}
   \label{Fig2}
\end{figure}

Our kinematical notations are as follows ($m$ is the nucleon  mass and we put all meson masses to zero):
\begin{eqnarray}
&&q=k-k'~~~~~; ~~~~~Q^2 = -q^2~~~~~; ~~~~~\Delta = p_2-p_1 ~~~~~; ~~~~~\Delta^2=t \,;\nonumber\\
&&p_1^\mu=(1+\xi)p^\mu +\frac{1}{2}  \frac{m^2-\Delta_T^2/4}{1+\xi} n^\mu -\frac{\Delta_T^\mu}{2}~~~~;~~~~ p_2^\mu=(1-\xi)p^\mu +\frac{1}{2}  \frac{m^2-\Delta_T^2/4}{1-\xi} n^\mu +\frac{\Delta_T^\mu}{2}\,; \\
&&q^\mu= -2\xi' p^\mu +\frac{Q^2}{4\xi'} n^\mu ~;~\epsilon_L^\mu(q)= \frac{1}{Q} [2\xi' p^\mu +\frac{Q^2}{4\xi'} n^\mu] ~;~p_M^\mu=  2(\xi-\xi') p^\mu +\frac{-\Delta_T^2}{4(\xi-\xi')}  n^\mu -\Delta_T^\mu \,,\nonumber
\end{eqnarray}
with $p^2 = n^2 = 0$ and $p.n=1$. The skewness variable $\xi$ is defined as
 \begin{eqnarray}
\xi = - \frac{(p_2-p_1).n}{2} = - \frac{q.n}{2}  \,.
\end{eqnarray} 
Neglecting the nucleon mass and $\Delta_T$, the approximate values of $\xi$ is
\begin{eqnarray}
\xi \approx \frac{Q^2}{4p_1.q-Q^2}  = \frac{x_B }{2-x_B }\,.
\end{eqnarray}
with
$ x_B \equiv \frac {Q^2}{2p_1.q}\,.$

\section{Distribution amplitudes and GPDs} 
In the collinear factorization framework, the hadronization of the quark-antiquark pair is described by a distribution amplitude(DA) which obeys a twist expansion and evolution equations. At leading twist, it reads for the pseudoscalar case :
\begin{eqnarray}
\langle M^+(P_M) | \bar u_(y) \gamma^5 \gamma^\mu d(-y) |0 \rangle & =&
   i f_MP_M^\mu \int_0^1 dz e^{i(z-\bar z)P_M.y}  \phi_M(z)\,,
         \end{eqnarray}
         and for the longitudinal vector meson case:
         \begin{eqnarray}
\langle M_L^+(P_M,\varepsilon_L) | \bar u(y) \gamma^\mu d(-y) |0 \rangle & =&
    f_M P_M^\mu \int_0^1 dz e^{i(z-\bar z)P_M.y}  \phi_M(z)\,,
         \end{eqnarray}
with $z=\frac{k^+}{P_M^+}$ and  where $\int_0^1 dz ~ \phi_M(z) = 1$,  $f_\pi= 0.131$ GeV and $f_\rho= 0.216$ GeV. As usual, we denote $\bar z=1-z$. 
We define the gluon and quark generalized parton distributions of a parton $q$ (here $q = u,\ d$) in the nucleon target   with the conventions of \cite{MD}.

 \section{The scattering amplitude}
  When there is a change in the baryonic flavor, as in the reaction $\nu_l (k)n(p_1) \to l^- (k')M^0 (p_M)p'(p_2) $, the amplitude does not depend on gluon GPDs. In the other cases, namely $\nu_l (k)p(p_1) \to l^- (k')M^+ (p_M)p'(p_2) $ and $\nu_l (k)n(p_1) \to l^- (k')M^+ (p_M)n'(p_2) $, there is a gluonic contribution coming from the diagrams of Fig.2.

 \subsection{The quark contribution}
 The flavor sensitive electroweak vertex selects the $d \to u$ and the $\bar u  \to \bar d$ transitions.   Moreover, reactions such as $\nu n \to l^- M^0 p$ give access to the neutron $\to$ proton GPDs ($H^{du}, E^{du}$...) which are related to the differences of GPDs through $H^{du}(x,\xi,t)=H^d(x,\xi,t)-H^u(x,\xi,t)$.
 The four Feynman diagrams of Fig. 1 contribute to the coefficient function, the diagrams (a,b) are attached to the $d-$quark GPDs while the  
diagrams (c,d) are attached to the $u-$quark GPDs. 
The chiral-odd GPDs do not contribute to the longitudinal amplitude since the coefficient function does not depend on any transverse vector. 

The vector and axial hard amplitudes (without the coupling constants) read:
\begin{eqnarray}
{\cal M}_{ab}^V &=&   \left\{ \frac{Tr_a}{D^q_1 D^q_2} + \frac{Tr_b}{D^q_1 D^q_3}  \right\} \,, ~~~
{\cal M}_{cd}^V =   \left\{ \frac{Tr_c}{D^q_4 D^q_5} + \frac{Tr_c}{D^q_4 D^q_6}  \right\} \,,\\
{\cal M}_{ab}^5 &=&  \left\{ \frac{Tr_a^5}{D^q_1 D^q_2} + \frac{Tr_b^5}{D^q_1 D^q_3} \right\} \,, ~~~
{\cal M}_{cd}^5 =  \left\{ \frac{Tr_c^5}{D^q_4 D^q_5} + \frac{Tr_d^5}{D^q_4 D^q_6} \right\} \,,
\end{eqnarray}
where the propagators  are :
\begin{eqnarray}
D^q_1 = [ (x-\xi)p+ \bar z p_M ]^2+i\epsilon  = \bar z (x-\xi) \frac{Q^2}{2\xi }+i\epsilon\,&,& D^q_4 =  z (-x-\xi) \frac{Q^2}{2\xi }+i\epsilon\,,\nonumber\\
D^q_2 = [ (x+\xi)p+q ]^2+i\epsilon  =  \frac{Q^2}{2\xi} (x-\xi+i\epsilon)\,&,& D^q_5 =   \frac{Q^2}{2\xi} (-x-\xi+i\epsilon)\,,\\
D^q_3 = [ 2 \xi p- \bar z p_M ]^2+i\epsilon  = -\bar z Q^2+i\epsilon\,= -\bar z Q^2 \,&,& D^q_6 = - z Q^2\, ,\nonumber
\end{eqnarray}
and the traces are :
\begin{eqnarray}
Tr_a = Tr_a^5= -\frac{Q^3}{\xi^2 } (x-\xi) \,~~~&,&~~~
Tr_b = Tr_b^5 =  2\bar z  \frac{Q^3}{\xi}\,, \nonumber \\
Tr_c = Tr_c^5= -\frac{Q^3}{\xi^2 } (x+\xi) \,~~~&,&~~~
Tr_d = Tr_d^5 =  -2 z  \frac{Q^3}{\xi}\,.
\end{eqnarray}
Using the symmetry of the meson DA with respect to $(z \leftrightarrow \bar z)$, and the fact that  diagrams (a,b) and diagrams (c,d) are related by a substitution $(z, x \leftrightarrow \bar z, -x)$ , one may then write 
the quark contribution to the amplitude as a convolution of chiral-even GPDs 
\begin{equation}
 H^{\nu}(x,\xi,t) = H^d(x,\xi,t) - H^u(-x,\xi,t)
 \label{Fnu}
 \end{equation}
  (and similar definitions for $ \tilde H^{\nu}(x,\xi,t),  E^{\nu}(x,\xi,t)$ and $\tilde  E^{\nu}(x,\xi,t)$ ) as:
\begin{eqnarray}
T^q  &=& \frac{- iC_q}{2Q} \bar{N}(p_{2}) \left[  {\mathcal{H}}^{\nu} \hat n - \tilde {\mathcal{H}} ^{\nu}\hat n \gamma^5 
 + {\mathcal E}^{\nu} \frac{i\sigma^{n\Delta}}{2m_N} - \tilde {\mathcal E}^{\nu}\frac{\gamma ^5 \Delta.n}{2m_N}\right] N(p_{1}),
 \label{Tq}
\end{eqnarray}
with the chiral-even form factors defined as
\begin{eqnarray}
{\cal F }^{\nu} &=&2 f_{M}\int \frac{\phi_M(z)dz}{\bar z}\hspace{-.1cm}\int dx \frac{F^{\nu}(x,\xi,t)}{x-\xi +i\epsilon} ~~~~~ {\rm(pseudoscalar~meson)} ,\nonumber \\
{\cal F }^{\nu} &=&- 2 i f_{M}\int \frac{\phi_M(z)dz}{\bar z}\hspace{-.1cm}\int dx \frac{F^{\nu}(x,\xi,t)}{x-\xi +i\epsilon}  ~~~~~{\rm(vector ~meson)} ,
\label{LFF}
 \end{eqnarray} 
 for any chiral even quark  GPD in the nucleon $F(x,\xi,t)$.

  \subsection{The gluonic contribution}
Note that contrarily to the case of electroproduction of light pseudoscalar mesons, there is no $C-$parity argument to cancel the gluon contribution for the neutrino production of a pseudoscalar  meson. The six Feynman diagrams of Fig. 2 contribute to the coefficient function when there is no charge exchange between the nucleons. The three last ones correspond to the tree first ones with the substitution $x\leftrightarrow -x$, and an overall minus sign for the axial case.
The transversity gluon GPDs do not contribute to the longitudinal amplitude since there there is no way to flip the helicity by two units when producing a (pseudo)scalar meson.

 The symmetric and antisymmetric hard amplitudes read:
\begin{eqnarray}
g_\perp^{ij} {\cal M}_H^S &=&   \left\{ \frac{Tr_a^S}{D_1 D_2} + \frac{Tr_b^S}{D_3 D_4} + \frac{Tr_c^S}{D_4 D_5}\right\}  + \left\{x\rightarrow  -x\right\} \\
i \epsilon_\perp^{ij}{\cal M}_H^{A} &=&  \left\{ \frac{Tr_a^A}{D_1 D_2} + \frac{Tr_b^A}{D_3 D_4} + \frac{Tr_c^A}{D_4 D_5} \right\}- \left\{x\rightarrow  -x\right\}
\end{eqnarray}
where the traces are:
\begin{eqnarray}
Tr_a^S =  \frac{2 \bar z }{Q} g_T^{ij}\left[\frac{\xi-x}{2\xi} Q^4\right]\,&,&
Tr_a^A =  \frac{2i \bar z \epsilon^{npij}}{Q} \left[ \frac{x-\xi}{2\xi} Q^4\right]\,,\\
Tr_b^S =   \frac{2z }{Q} g_T^{ij}\left[  \frac{\xi -x}{2\xi}Q^4  \right]\,&,&
Tr_b^A =  \frac{2iz  \epsilon^{npij}}{Q} \left[\frac{\xi -x}{2\xi} Q^4 \right] \,,\\
Tr_c^S =  \frac{- 1}{\xi Q}g_T^{ij} \left[  \frac{x^2-\xi^2}{2\xi} Q^4   \right] \,&,&
Tr_c^A =   \frac{- i  \epsilon^{npij}}{\xi Q} \left[\frac{x^2- \xi^2}{2\xi}Q^4\right]\,,
\end{eqnarray}
and the denominators read
\begin{eqnarray}
D_1 &=& - \bar z Q^2 +i \epsilon \,, D_2 =  \bar z \frac{Q^2}{2\xi} (x-\xi +i\epsilon)\,,D_3 = -z Q^2 +i \epsilon  \,,\\
D_4 &=& z \frac{Q^2}{2\xi} (x-\xi +i\epsilon)\, ,
D_5 = \  \frac{\bar z Q^2}{2\xi} [ -x-\xi +i \epsilon]  \,.
\end{eqnarray}
The axial amplitude vanishes after summing the diagrams. The gluonic contribution to the amplitude thus reads:
\begin{eqnarray}
T^g &=&  \frac{ i C_g}{2} \int_{-1}^{1}dx \frac{- 1}{(x+\xi-i\epsilon)(x-\xi+i\epsilon)} \int_0^1 dz f_M \phi_M(z)  \bar{N}(p_{2})[H^g\hat n +E^g\frac{i\sigma^{n\Delta}}{2m} ]N(p_{1}) {\cal M}_H^S \, \\
&\equiv&  \frac{- i C_g}{2Q}  \bar{N}(p_{2}) \left[ {\cal H}^g\hat n +{\cal E}^g\frac{i\sigma^{n\Delta}}{2m} \right] N(p_{1}) \,, 
\end{eqnarray}
with  $C_g= T_f\frac{\pi}{3} \alpha_s V_{du}$, $T_f=\frac{1}{2}$ and the factor $\frac{- 1}{(x+\xi-i\epsilon)(x-\xi+i\epsilon)}$ comes from the conversion of the gluon field to the strength tensor. The  gluonic form factors ${\cal H}^g$,  ${\cal E}^g$ read
\begin{eqnarray}
{\cal F }^{g} &=&\frac{8 f_{M}}{\xi} \int _0^1\frac{\phi(z)dz}{z \bar z} \int _{-1}^{1}dx \frac{F^{g}(x,\xi,t)}{x-\xi +i\epsilon} \,~~~~~ {\rm (pseudoscalar ~meson)}\,,\nonumber \\
{\cal F }^{g} &=&\frac{- 8 i f_{M}}{\xi} \int _0^1\frac{\phi(z)dz}{z \bar z} \int _{-1}^{1}dx \frac{F^{g}(x,\xi,t)}{x-\xi +i\epsilon} \,  ~~~~~ {\rm (vector ~meson)} .
\label{GFF}
 \end{eqnarray} 
Note that there is no singularity in the integral over $z$ if the DA vanishes like $z \bar z$ at the limits of integration.

 \section{Cross sections}
  The differential cross section for neutrino production of a meson is written as:
 \begin{eqnarray}
\label{cs}
\frac{d^4\sigma(\nu N\to l^- N' M)}{dy\, dQ^2\, dt\,  d\varphi}
%&=&
%-\varepsilon\cos(2\varphi)\sigma_{+-}
%\\
 = \bar\Gamma \varepsilon\sigma_{L} \,,
\end{eqnarray}
with $y= \frac{p \cdot q}{p\cdot k}$ , $Q^2 = x_B y (s-m^2)$, $\varepsilon \approx \frac{1-y}{1-y+y^2/2}$ and
\begin{equation}
\bar \Gamma = \frac{G_F^2}{(2 \pi)^4} \frac{1}{32y} \frac{1}{\sqrt{ 1+4x_B^2m_N^2/Q^2}}\frac{1}{(s-m_N^2)^2} \frac{Q^2}{1-\epsilon}\,, \nonumber
\end{equation}
%FACTOR TWO CORRECTED
where the longitudinal cross-section $\sigma_{L}=\epsilon_L^{* \mu} W_{\mu \nu} \epsilon_L^\nu$ is the product of  amplitudes for the process $ W(\epsilon_L) N\to M N' $, averaged  (summed) over the initial (final) nucleon polarizations.  $\sigma_{L}$ is straightforwardly obtained by squaring the sum of the amplitudes $T^q+ T^g$; at zeroth order in $\Delta_T$, it reads  :
\begin{eqnarray}
\sigma_{L} =    \frac{1} { Q^2}\biggl\{ && [\, |C_q{\mathcal{H}}^{\bar q} + C_g{\mathcal{H}}_{g}|^2 + |C_q\tilde{\mathcal{H}}^{\bar q}|^2 ] (1-\xi^2) +\frac{\xi^4}{1-\xi^2} [\,   |C_q {\mathcal{E}}^{\bar q}+C_g {\mathcal{E}}_{g} |^2 + |C_q\tilde{\mathcal{E}}^{\bar q} |^2 ]  \nonumber  \\
  &&  -2 \xi^2 {\mathcal R}e  [C_q{\mathcal{H}}^{\bar q} + C_g{\mathcal{H}}_{g}] [C_q {\mathcal{E}}^{\bar q }+C_g {\mathcal{E}}_{g}] ^* -2 \xi^2 {\mathcal R}e  [C_q\tilde{\mathcal{H}}^{\bar q} ] [C_q \tilde{\mathcal{E}}^{\bar q }]^* \biggr\} .\, 
\end{eqnarray}

\begin{figure}
\includegraphics[width=0.8\textwidth]{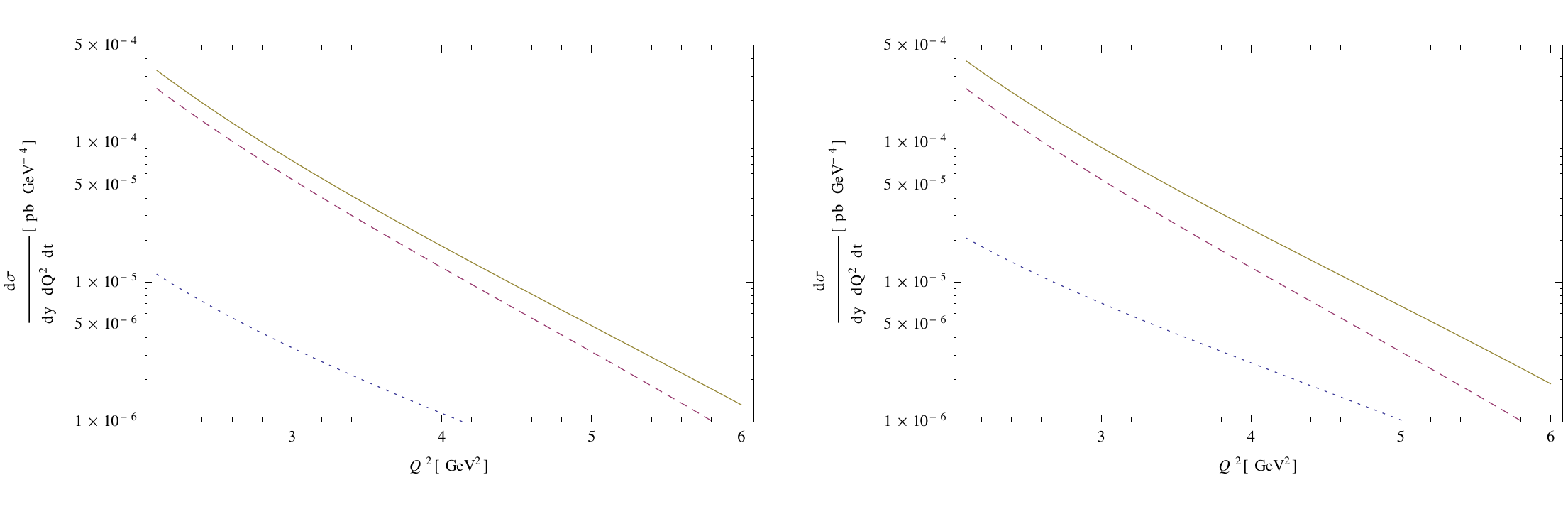}
%\vspace{1cm}
\caption{The $Q^2$ dependence of the  cross section $\frac{d^3\sigma(\nu N \to l^- N \pi^+)}{dy\, dQ^2\, dt}$ (in pb GeV$^{-4}$) for $y=0.7, \Delta_T = 0$  and $s=20$ GeV$^2$, on a proton (left panel) and on a neutron (right panel). The quark contribution (dotted curves) is significantly smaller than the gluon contribution (dashed curves). The solid curves are the sum of the (quark + gluon+ interference) contributions. }
   \label{sigma_dy_piplus}
\end{figure}

\begin{figure}
\includegraphics[width=0.4\textwidth]{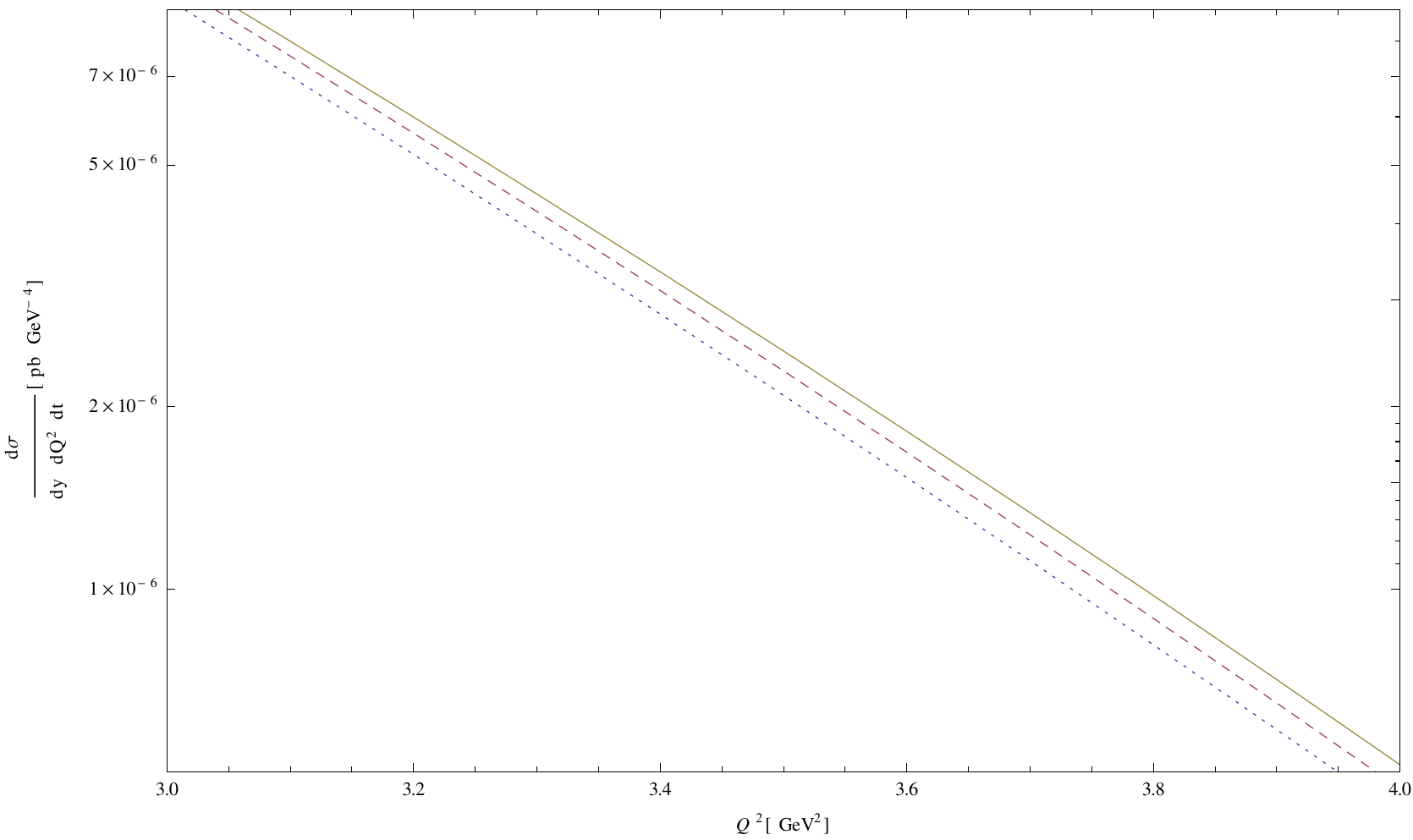} ~~~~~\includegraphics[width=0.36\textwidth]{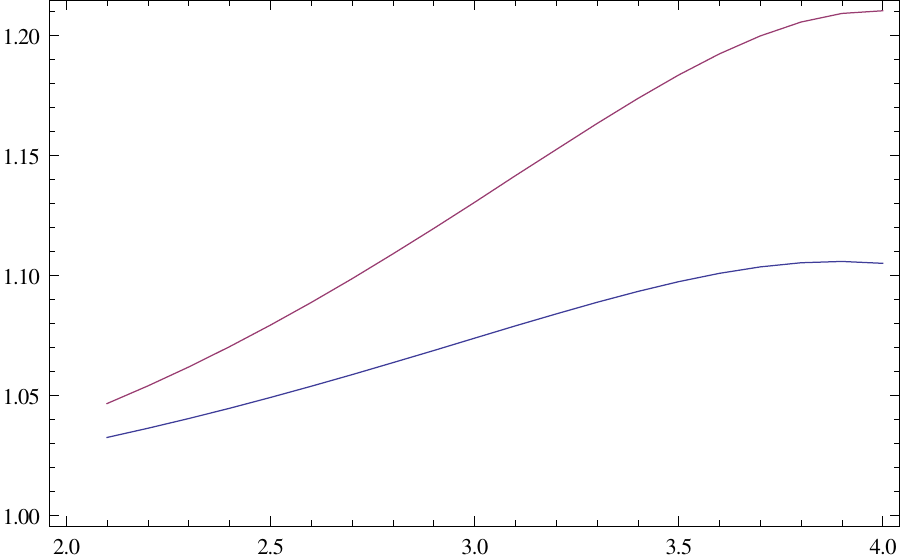}
%\vspace{1cm}
\caption{The $Q^2$ dependence of the  cross section $\frac{d^3\sigma(\nu N \to l^- N \pi^+)}{dy\, dQ^2\, dt}$ (in pb GeV$^{-4}$) for $y=0.5, \Delta_T = 0$  and $s=13$ GeV$^2$, on a proton (left panel). The dotted line corresponds to  $ E^g (x, \xi, t)= 0$, the dashed (resp. solid) one uses V2 (resp. V4) parametrization of  $ E^g (x, \xi, t)$ from Ref. \cite{Metz}.  On the right panel we show the ratio of the cross section calculated with V4 of $E_g$ (upper line) and V2 of $E_g$ (lower line) to the cross section calculated with $E_g = 0$.}
   \label{PiPlus_Eg_V2_V4}
\end{figure}

\begin{figure}
\includegraphics[width=0.9\textwidth]{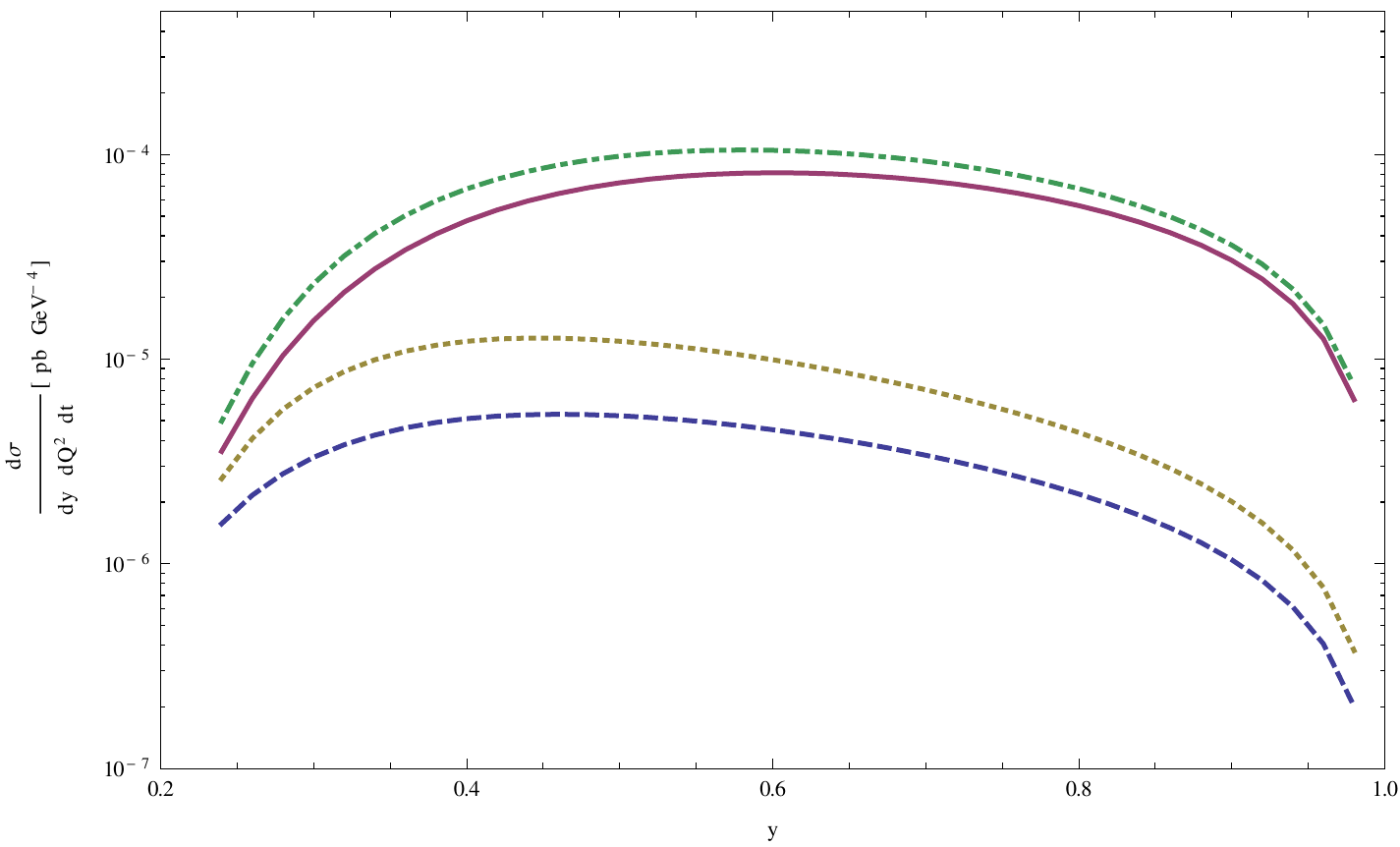}
%\vspace{1cm}
\caption{The $y$ dependence of the  cross section $\frac{d^3\sigma(\nu N \to l^- N \pi^+)}{dy\, dQ^2\, dt}$ (in pb GeV$^{-4}$) for $Q^2= 3$ GeV$^2$, $ \Delta_T = 0$  and $s=20$ GeV$^2$, on a proton (solid) and on a neutron (dot-dashed). The lower  curves are the quark contributions for the proton (dashed) and the neutron(dotted) target.}
   \label{ydep}
\end{figure}

\subsection{$\nu N \to l^- \pi^+ N$}

Let us now estimate various cross sections which may be accessed with a neutrino beam on a nucleus. Firstly, 
\begin{eqnarray}
\nu_l (k)p(p_1) &\to& l^- (k') \pi^+ (p_M)p'(p_2) \,,\\
\nu_l (k)n(p_1) &\to& l^- (k')\pi^+ (p_M)n'(p_2) \,,
\end{eqnarray}
allow both quark and gluon GPDs to contribute.

The relative importance of quark and gluon contributions to the  cross sections are shown on Fig. \ref{sigma_dy_piplus} as a function of $Q^2$ for a specific set of kinematical variables. The gluon contribution strongly dominates the cross-section. 
In Fig. \ref{PiPlus_Eg_V2_V4}, we show the relative importance of the $H^g$ and $E^g$ contributions, using two models for the latter GPD, as described in \cite{Metz}.
The $y$ dependence is displayed on Fig. \ref{ydep}. The  cross section vanishes as $y\to 1$ as is obvious from Eq.(\ref{cs}).

Our estimates are obtained by the use of the pion asymptotic DA. The studies which attempt to take into account non-perturbative effects (as those based on Schwinger-Dyson equation \cite{Chang:2013nia}  or light-front quark model \cite{Choi:2007yu})  resulting in nonvanishing quark masses and condensates lead to pion DAs which forms differ significantly from the form of the asymptotic DA, see e.g. Fig.1 and references of  \cite{Mikhailov:2016lof}.
Our predictions obtained with such non-asymptotic pion DAs give  values of cross sections which are easily obtained from the asymptotic ones by multiplying by the factor $  |  K  | ^2$ with 
\begin{eqnarray}
K = \frac {\int dz \phi(z)/z } {\int dz \phi_{as}(z)/z }\, .
\end{eqnarray}

We get  $  |  K  | ^2 \approx  3.2$  for the DA of \cite{Chang:2013nia}, $  |  K  | ^2 \approx  1.2, 0.96$ for the two models described in  \cite{Choi:2007yu} and $  |  K  | ^2 \approx  1.8$ for the AdS-CFT prediction of Ref.\cite{Brodsky:2006uqa}. 

\subsection{$\nu N \to l^- \rho_L^+ N$}

The vector meson case 
\begin{eqnarray}
\nu_l (k)p(p_1) &\to& l^- (k') \rho_L^+ (p_M)p'(p_2) \,,\\
\nu_l (k)n(p_1) &\to& l^- (k')\rho_L^+ (p_M)n'(p_2) \,,\\
\end{eqnarray}
is identical to the pseudoscalar one, with the obvious change $f_\pi \to f_\rho$, which yields a multiplicative  factor $2.72$ at the cross section level, to be modified if the distribution amplitudes for vector mesons have different shapes from the pseudoscalar ones. The $Q^2$ and $y$ dependences of the cross sections may thus be read off Figs. \ref{sigma_dy_piplus}, \ref{PiPlus_Eg_V2_V4} and \ref{ydep}.

\subsection{$\nu n \to l^- \pi^0 p$ and $\nu n \to l^- \rho^0 p$}
\begin{figure}
\includegraphics[width=0.8\textwidth]{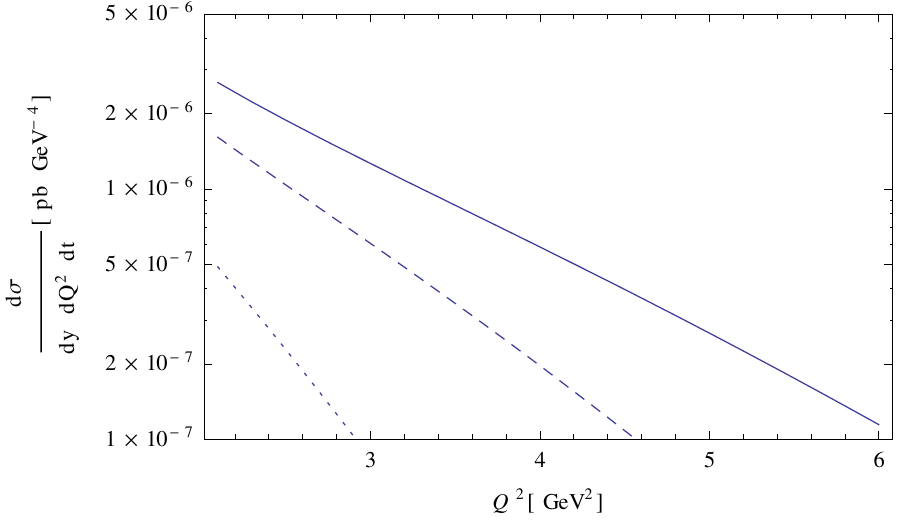}
%\vspace{1cm}
\caption{The $Q^2$ dependence of the  cross section $\frac{d^3\sigma(\nu n \to l^- p \pi^0)}{dy\, dQ^2\, dt}$ (in pb GeV$^{-4}$) for $ \Delta_T = 0$  and $s=20$ GeV$^2$.   The solid (resp. dashed, resp. dotted) line corresponds to $y=0.7$ (resp. $ 0.5$, resp. $0.3$). There is no gluon contribution to this amplitude. }
   \label{sigma_dy_pizero}
\end{figure}

The reaction
\begin{eqnarray}
\nu_l (k)n(p_1) &\to& l^- (k')M^0 (p_D)p(p_2) \,, 
\end{eqnarray}
does not benefit from gluon GPDs contributions, but only from the flavor changing $F^{d\to u}_{n\to p} (x, \xi, t) =F^u_p (x, \xi, t)-F^d_p(x, \xi, t)$ GPDs ($F$ denotes here any GPD). The analogue of Eq. \ref{Tq} for the amplitude for $\pi^0$ production is obtained by changing in the numerator:
\begin{eqnarray}
H_d(x,\xi)-H_u(-x,\xi)) \to \frac{1}{\sqrt2} [ (H_u(x,\xi)-H_d(x,\xi)) + (H_u(-x,\xi)-H_d(-x,\xi))].
\end{eqnarray}
We show on Fig.\ref{sigma_dy_pizero} the differential cross section for the production of a $\pi^0$ at $\Delta_T = 0, s= 20$ GeV$^2$ and $y=0.7$ (resp. $ 0.5$, resp. $0.3$) as a solid (resp. dashed, resp. dotted) line . The cross-section is rather small due to the absence of the gluon contribution. The  leading twist $\rho_L^0$ production cross section is proportional to this cross section; as above, the same discussion of the dependence on the distribution amplitudes of the mesons   leads to model-dependent  proportionality factors.

\subsection{Antineutrino cross sections :  $\bar \nu p \to l^+ \bar M^0 n$ and $\bar \nu N \to l^+ \bar M^- N$}
Let us now present the results for the antineutrino case. Although smaller than the neutrino flux, the antineutrino flux is always sizable, as discussed recently in \cite{Ren:2017xov}.
\begin{eqnarray}
 \bar\nu_l (k) p(p_1) &\to& l^+ (k') M^-(p_M) p' (p_2)\,,\\
  \bar\nu_l (k) p(p_1) &\to& l^+ (k') M^0(p_M) n' (p_2)\,,\\
  \bar\nu_l (k) n(p_1) &\to& l^+ (k') M^-(p_M) n' (p_2)\,,
\end{eqnarray}

  Going from the neutrino to the antineutrino case essentially leads to a transformation $z \to \bar z$ and $x \to -x$ in the expression of the amplitude. Using the fact that $\phi^{M^-}(\bar z) = \phi^{M^+}( z)$, the amplitudes can be written in terms of the same DA, but taking the GPD as $H(-x, \xi, t), \tilde H(-x, \xi, t), ...$. For obvious reasons, the gluon contributions are the same as for the neutrino case and, taking into account the structure of the GPD combination $F^\nu (x,\xi,t) $ in Eq. [\ref{Fnu}] the quark contributions are also equal, so that the antineutrino cross sections on a proton (resp. neutron) are identical to the neutrino cross sections on a neutron (resp. proton): 
\begin{eqnarray}  
 d\sigma (\bar \nu_l  ~p \to l^+  M^- p' )  &=&  d\sigma (\nu_l  ~n \to l^-  M^+ n' )  \, , \\
  d\sigma (\bar \nu_l  ~p \to l^+  M^0 n' )  &=&  d\sigma (\nu_l  ~n \to l^-  M^0 ~p' )  \, , \\
   d\sigma (\bar \nu_l  ~n \to l^+  M^- n' )  &=&  d\sigma (\nu_l  ~p \to l^-  M^+ p' )  \, .
 \end{eqnarray}
The  differential cross sections for the anti-neutrino production can thus be read from the neutrino case shown in Figs. (3-6). 

\section{Conclusions}
 Collinear QCD factorization allows us  to calculate exclusive neutrino production of mesons in terms of quark and gluon GPDs for large enough $Q^2$ (for a purely hadronic description of low $Q^2$ kinematics, see for instance Ref. \cite{Hernandez:2016yfb}). We complemented the previous calculations \cite{Kopeliovich:2012dr} which were omitting gluon contributions.
 We have demonstrated that gluon and  chiral-even quark GPDs contribute in specific ways to the amplitudes.  The flavor dependence, and in particular the difference between $M^+$ and $M^0$ production rates, allows to test the importance of gluonic contributions, which we predict to be large in the first case. The behaviour of the proton and neutron target cross sections enables to separate the $u$ and $d$ quark contributions. Those properties may be very useful for future GPDs extractions programs e.g. \cite{PARTONS}.

An interesting observation is that neutrino production of $\pi$ mesons and longitudinally polarized $\rho$ mesons are proportional at leading twist (and all orders in $\alpha_S$). This is very different from the electroproduction case, where different GPDs contribute to the pseudoscalar and vector meson production amplitudes. Since the phenomenology of these two cases turned out to be very controversial  \cite{pirho}, in particular with respect to the dominance of leading twist pseudoscalar meson distribution amplitudes, the study of the neutrino case should be very informative to disentangle the role of nucleon GPDs and of meson DAs in the apparent breaking of leading twist dominance at moderate scales.

Some remarks are in order about possible extensions of the present study of neutrino production of a pion.
First let us note that in the similar framework we can study neutrino production of  two pions with a small (e.g. $< 1 $GeV) invariant mass. Such a pion pair is described within the collinear factorisation of QCD by a non pertubative generalized distribution amplitudes (GDAs)  \cite{GDA}.  Since two produced pions can be in different isospin states
the GDAs can be constrained by measurements of observables which are sensitive to interference between isospin states of two pions \cite{DI}.

One can also extend the analysis of the present paper from the study of neutrino production of a pion in  forward kinematics  to the similar reaction in  backward
kinematics. The theoretical description of this last case involves the non perturbative transition distribution amplitudes \cite{TDA} for nucleon to pion transition  and can be performed in analogy with the studies of Ref. \cite{TDA2}.

We did not discuss the role of nuclear effects in deep exclusive neutrino production, although this question is central in many experiments, as recently reviewed in  \cite{revnuc}. More work is obviously needed in this direction, both for light nuclei and heavy nuclei \cite{nuclearGPD}. Let us briefly mention that contrarily to the  case of the total cross section, one expects the phenomenon of color transparency \cite{CT} to decrease these nuclear effects when $Q^2$ grows. Indeed,  in a hard exclusive process such as the one studied here,  the meson is produced with a small transverse size which grows while it propagates through the nucleus, leading to a smaller effective final state interaction cross section. 

 Planned medium and high energy neutrino facilities \cite{NOVA} and experiments such  as Miner$\nu$a \cite{Aliaga:2013uqz} and MINOS+ \cite{Timmons:2015laq} will thus allow  some important progress in the realm of hadronic physics.

%%%%%%%%

%%%%%%%%%%%%%%%%%%%%%%%%%%%%%%%%%%%%%%%%%%%%%%%%%%%%%%%%%%%%%%%%%%%%%%%%%%% 
\paragraph*{Acknowledgements.}
\noindent
 This work is partly supported by grant No 2015/17/B/ST2/01838 from the National Science Center in Poland, by the Polish-French collaboration agreements  Polonium and COPIN-IN2P3, and  by the French grant ANR PARTONS (Grant No. ANR-12-MONU-0008-01).

\end{document}